\documentclass[a4paper]{aipproc}

\layoutstyle{6x9}

\begin{document}

\title{Scaling of Anisotropic Flows in Intermediate Energy and Ultra-relativistic Heavy Ion
Collisions}
\author{ Y. G. Ma}{
      address={Shanghai Institute of Applied Physics,
     Chinese Academy of Sciences,   Shanghai 201800, China}
     }

\begin{abstract}
Anisotropic flows ($v_2$ and $v_4$)  of hadrons and light nuclear
clusters are studied by a partonic transport model and nucleonic
transport model, respectively, in ultra-relativistic and
intermediate energy heavy ion collisions. Both
number-of-constituent-quark scaling of hadrons, especially for
$\phi$ meson which is composed of strange quarks,  and
number-of-nucleon scaling of light nuclear clusters are discussed
and explored for the elliptic flow ($v_2$). The ratios of
$v_4/v_2^2$ of hadrons and nuclear clusters are, respectively,
calculated and they show different constant values which are
independent of transverse momentum. The above phenomena can be
understood, respectively, by the coalescence mechanism in
quark-level or nucleon-level.
\end{abstract}
\date{\today}

\maketitle

\section{Introduction}

Anisotropic   flow is of interesting subject in  theoretical and
experimental investigations on  heavy ion collision dynamics
\cite{Olli,Voloshin,Sorge,Danile,Zhang,Shu,Kolb,Zheng,Gale,INDRA}.
Many studies of  the 1-th and 2-nd anisotropic flows, namely the
directed flow and elliptic flow, respectively, revealed much
interesting physics about the properties and origin of the
collective flows. In particular, recent ultra-relativistic Au + Au
collision experiments demonstrated the number of constituent-quark
(NCQ) scaling from the transverse momentum dependence of the
elliptic flow for  different mesons and baryons at the
Relativistic Heavy Ion Collider (RHIC) in Brookhaven National
Laboratory (BNL) \cite{J.Adams}, it indicates that the partonic
degree of the freedom  plays a dominant role in formation of the
dense matter in the early stage of collisions. Several theoretical
models were proposed to interpret the NCQ-scaling of hadrons at
RHIC \cite{Ko,Molnari,Duke,Hwa,Chen}. Of which, a popular
interpretation is assuming that mesons and baryons are formed by
the coalescence or recombination of the constituent quarks. While
at intermediate energy heavy ion collisions, the coalescence
mechanism has been also used to explain the formation of light
fragments  and their spectra of kinetic energy or momentum of
light particles some times ago
\cite{Awes,Mekjian,Sato,Llope,Hagel}. A few studies investigated
the mass dependence of the directed flow \cite{Huang,Kunde}.
However, systematic theoretical studies on the elliptic flow of
different nuclear fragments in intermediate energy domain in terms
of the coalescence mechanism is rare.

In this work,  we will investigate the  flow scaling of hadrons
and of nuclear fragments with different microscopic transport
theories, namely the partronic transport model in
ultra-relativistic energies  and nucleonic transport model at
intermediate energies, respectively. A common naive coalescence
mechanism in quark level or nucleonic level was applied to analyze
the flow. Hadrons, namely baryons and mesons, can be formed by the
quark-level coalescence. While, nuclear fragments can be formed by
the nucleon-level coalescence. Dependences of the
number-of-constituent-quark or number-of-nucleon (NN) of the
elliptic flow are surveyed, respectively, and the ratio of $v_4$
and $v_2^2$ is studied.

Two examples are given to illustrate the NCQ or NN scaling of the
elliptic flow in ultra-relativistic and intermediate energy HIC,
respectively.  A Multi-Phase Transport (AMPT) model was applied to
simulate Au+Au at $\sqrt{s}$ = 200 GeV/c for the former case and
Isospin-Dependent Molecular Dynamics (IDQMD) was used to simulate
$^{86}$Kr + $^{124}$Sn at $E_{beam}/A$ = 25 MeV/u for the later
case.

\section{Definition of anisotropic flow and Model Introduction }

\subsection{ Definition of anisotropic flow}

Anisotropic flow is defined as the different $n$-th harmonic
coefficient $v_n$ of an azimuthal Fourier expansion of the
particle invariant distribution \cite{Voloshin}
\begin{equation}
\frac{dN}{d\phi} \propto {1 + 2\sum_{n=1}^\infty v_n cos(n\phi) },
\end{equation}
where $\phi$ is the azimuthal angle between the transverse
momentum of the particle and the reaction plane. Note that in the
coordinate system the $z$-axis along the beam axis, and the impact
parameter axis is labelled as $x$-axis.  The first harmonic
coefficient $v_1$ represents the directed flow,
$v_1 = \langle cos\phi \rangle = \langle \frac{p_x}{p_t} \rangle$,
where $p_t = \sqrt{p_x^2+p_y^2}$ is transverse momentum. $v_2$
represents the elliptic flow which characterizes the eccentricity
of the particle distribution in momentum space,
\begin{equation}
v_2 = \langle cos(2\phi) \rangle = \langle
\frac{p^2_x-p^2_y}{p^2_t} \rangle,
\end{equation}
and $v_4$ represents the 4-th momentum anisotropy,
\begin{equation}
v_{4} =\left\langle \frac{p_{x}^{4}-6p_{x}^{2}p_{y}^{2}+p_{y}^{4}}{%
p_{t}^{4}}\right\rangle . \label{v4}
\end{equation}

\subsection{ Partonic/hadronic transport model: AMPT model}
AMPT (A Multi-Phase Transport) model \cite{AMPT}  is a hybrid
model which consists of four main processes: the initial
condition, partonic interactions, the conversion from partonic
matter into hadronic matter and hadronic interactions. The initial
condition, which includes the spatial and momentum distributions
of minijet partons and soft string excitation, are obtained from
the HIJING model \cite{HIJING}. Excitation of strings  melt
strings into partons. Scatterings among partons are modelled by
Zhang's parton cascade model  (ZPC) \cite{ZPC}, which at present
includes only two-body scattering with cross section obtained from
the pQCD with screening mass. In the default version of the AMPT
model \cite{DAMPT} partons are recombined with their parent
strings when they stop interaction, and the resulting strings are
converted to hadrons by using a Lund string fragmentation model
\cite{Lund}. In the string melting version of the AMPT model (we
briefly call it as "the melting AMPT" model)\cite{SAMPT}, a simple
quark coalescence model is used to combine partons into hadrons.
Dynamics of the subsequent hadronic matter is then described by A
Relativistic Transport (ART) model \cite{ART}. Details of the AMPT
model can be found in a recent review \cite{AMPT}. It has been
shown that in previous studies \cite{SAMPT} the partonic effect
can not be neglected and the melting AMPT model is much more
appropriate than the default AMPT model when the energy density is
much higher than the critical density for the QCD phase transition
\cite{AMPT,SAMPT,Chen}. In the present work, the partonic
interacting cross section in AMPT model with string melting is
selected as 10mb.

\subsection{ Nucleonic transport model: QMD Model}

The Quantum Molecular Dynamics (QMD) approach is an n-body theory
to describe heavy ion reactions from intermediate energy to 2 A
GeV. It includes several important parts: the initialization of
the target and the projectile nucleons, the propagation of
nucleons in the effective potential, the collisions between the
nucleons, the Pauli blocking effect and the numerical tests. A
general review about QMD model can be found in \cite{Aichelin}.
The IDQMD model is based on QMD model affiliating the isospin
factors, which includes the mean field, two-body nucleon-nucleon
(NN) collisions and Pauli blocking \cite{Ma3,Liu,Wei,Ma2,Ma-hbt}.

In the QMD model each nucleon is represented by a Gaussian wave
packet with a width $\sqrt{L}$ (here $L$ = 2.16 ${\rm fm}^2$)
centered around the mean position $\vec{r_i}(t)$ and the mean
momentum $\vec{p_i}(t)$,
\begin{equation}
\psi_i(\vec{r},t) = \frac{1}{{(2\pi L)}^{3/4}}
exp[-\frac{{(\vec{r}- \vec{r_i}(t))}^2}{4L}] exp[-\frac{i\vec{r}
\cdot \vec{p_i}(t)}{\hbar}].
\end{equation}

The nucleons interact via nuclear mean field  and nucleon-nucleon
collision. The nuclear mean field can be parameterized by
\begin{equation}
U(\rho,\tau_{z}) = \alpha(\frac{\rho}{\rho_{0}}) +
\beta(\frac{\rho}{\rho_{0}})^{\gamma} +
\frac{1}{2}(1-\tau_{z})V_{c}  + C_{sym} \frac{(\rho_{n} -
\rho_{p})}{\rho_{0}}\tau_{z} + U^{Yuk}
\end{equation}
with $\rho_{0}$ the normal nuclear matter density (here, 0.16
$fm^{-3}$ is used). $\rho$, $\rho_{n}$ and $\rho_{p}$ are the
total, neutron and proton densities, respectively. $\tau_{z}$ is
$z$th component of the isospin degree of freedom, which equals 1
or -1 for neutrons or protons, respectively. The coefficients
$\alpha$, $\beta$ and $\gamma$ are parameters for nuclear equation
of state (EOS). $C_{sym}$ is the symmetry energy strength due to
the difference of neutron and proton. In the present work, we take
$\alpha$ = -124 MeV, $\beta$ = 70.5 MeV and $\gamma$ = 2.0 which
corresponds to the so-called hard EOS with an incompressibility of
$K$ = 380 MeV and $C_{sym}$ = 32 MeV \cite{Aichelin}. $V_{c}$ is
the Coulomb potential,  $U^{Yuk}$ is Yukawa (surface) potential.

The time evolution of the colliding system is given by the
generalized variational principal. Since the QMD can naturally
describe the fluctuation and correlation, we can study the nuclear
clusters in the model \cite{Aichelin,Ma3,Liu,Wei,Ma2,Ma-hbt}.
 In QMD model, nuclear clusters are usually recognized  by a simple
coalescence model: i.e. nucleons are considered to be part of a
cluster if in the end at least one other nucleon is closer than
$r_{min} \leq 3.5$ fm in coordinate space and $p_{min} \leq 300$
MeV/c  in momentum space  \cite{Aichelin}. This mechanism has been
extensively applied in transport theory for the cluster formation.

\section{Number-of-Constituent-Quark Scaling of the Elliptic Flow at RHIC Energies}

In non-central Au+Au collisions at RHIC energies, the overlap
region is anisotropic (nearly elliptic). Large pressure built up
in the collision center results in pressure gradient which is
dependent on azimuthal angle, which generates anisotropy in
momentum space, namely elliptic flow. Once the spatial anisotropy
disappears due to the anisotropic expansion, development of
elliptic flow also ceases. This kind of self-quenching process
happens quickly, therefore elliptic flow is primarily sensitive to
the early stage equation of state (EOS) \cite{elliptic-flow}.

\begin{figure}[htb]
\vspace*{-0.1in}
\centerline{\includegraphics[scale=0.6]{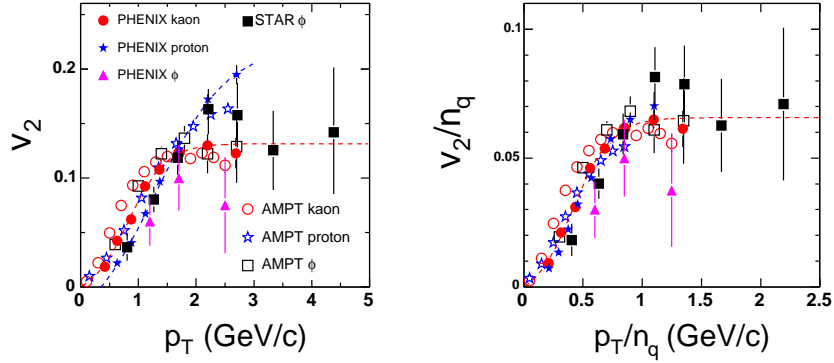}}
\vspace{-0.1cm}
 \caption{Left panel: $p_T$ dependence of $v_2$ for
$\phi$-meson which is compared with the cases of $K^{+}$+$K^{-}$
and $p+\overline{p}$. AMPT calculation are taken from
Ref.~\cite{Chen}.  Right panel: NCQ scaled $v_{2}$ as a function
of NCQ scaled transverse momentum. Lines represent NCQ-scaling
parameterization. \label{fig6}}
\end{figure}

Since many particles have been investigated already for
demonstrating NCQ scaling of elliptic flow at RHIC. However, the
study of strangeness particle, $\phi$, is rare. Figure~\ref{fig6}
shows both the measurement and the calculation  of $\phi$-meson
elliptic flow $v_{2}$ at RHIC \cite{CAI,Sarah,PHENIX-phi,Ma-UCLA}.
For comparison, $v_{2}$ of $K^{+}$+$K^{-}$ and $p+\overline{p}$
from \cite{PHENIX1} are plotted together. The most striking thing
in the figure is that the $\phi$-meson has a non-zero $v_{2}$ in
the hydrodynamic region ($p_{T} < 2$ GeV/$c$) and it has a
significant $v_{2}$ signal for $p_{T} > 2$ GeV/$c$ which is
comparable to that of the $K^{+}$+$K^{-}$. Since the formation of
$\phi$-meson through kaon coalescence at RHIC has been ruled out
by previous STAR measurement \cite{STAR-phi}, and the low
interaction cross-section of $\phi$-meson makes the contribution
to flow due to hadronic re-scattering processes very small
\cite{AMPT}, it may be possible to directly measure the flow of
strange quark via the flow of $\phi$-meson. Additionally, since
the $\phi$-meson $v_{2}$ values as a function of $p_{T}$ are
similar to those of other particles, this means that the heavier
$s$-quarks flow as much as the lighter $u$ and $d$ quarks. This
can happen if there are a significant number of interactions
between the quarks before hadronization. It is the signal of
partonic collectivity of the system! A prediction \cite{Chen} from
the naive coalescence model which is implemented in A Multi-Phase
Transport (AMPT) model with the string melting scenario
\cite{AMPT2} is also plotted here. The model describes the data
quite well at $p_{T} > 1.5$ GeV/c. This is another hint of
coalescence of strange quark as dominant production mechanism for
$\phi$-meson. Further evidence can be obtained from the
Number-of-Constituent-Quark (NCQ) picture in Figure~\ref{fig6}.

\begin{figure}[h]
  \resizebox{11pc}{!}{\includegraphics{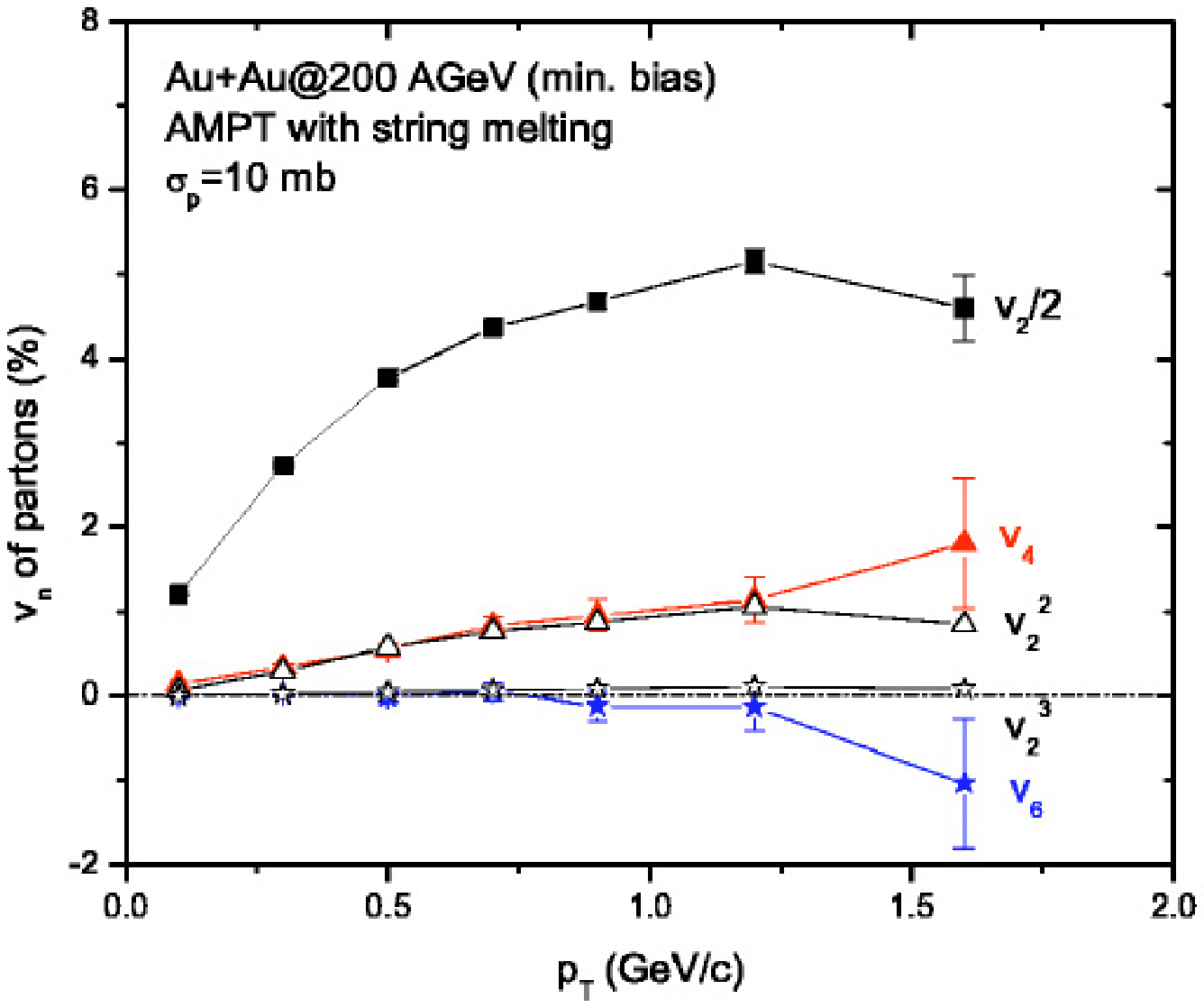}}
  \resizebox{12pc}{!}{\includegraphics{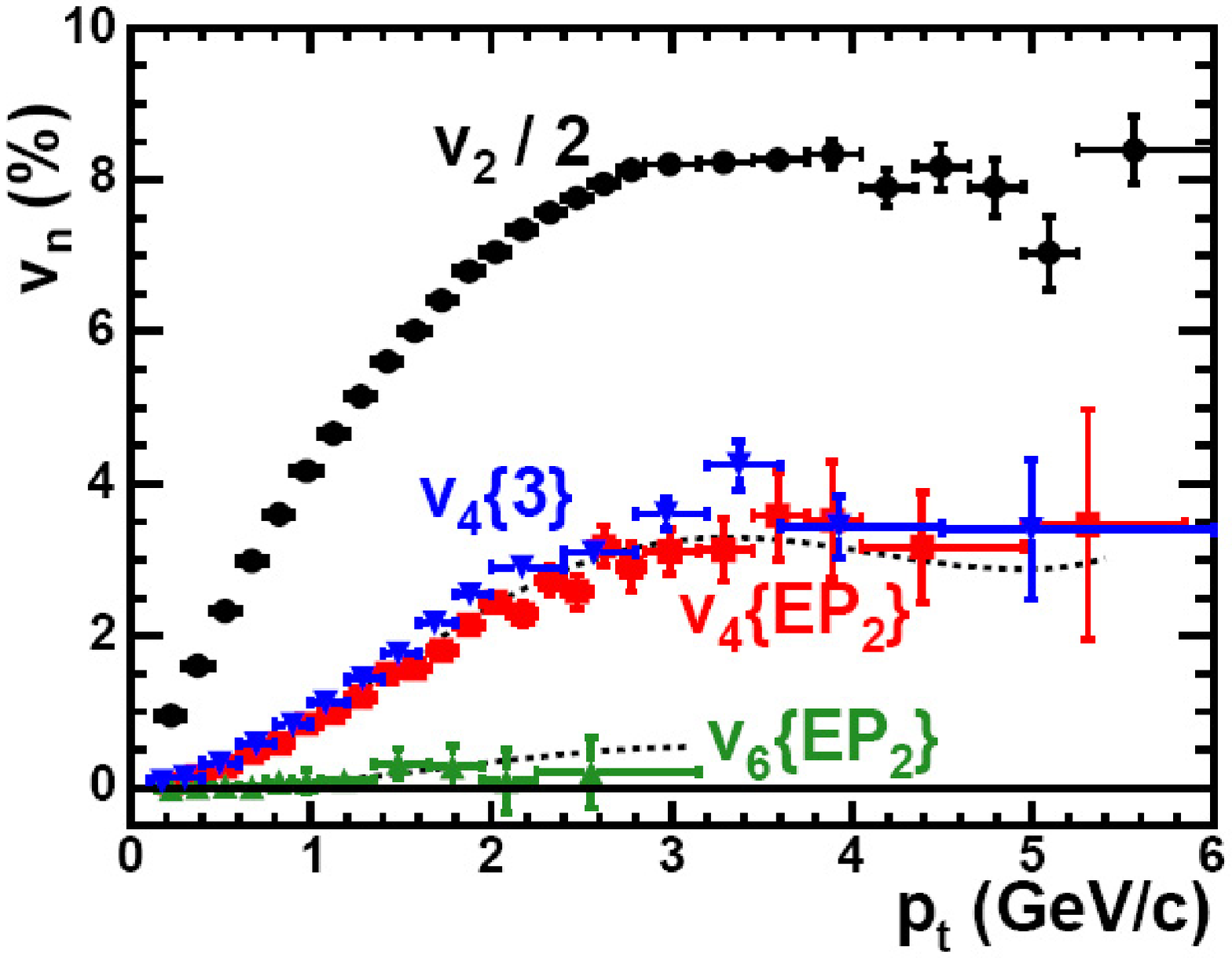}}
  \resizebox{13pc}{!}{\includegraphics{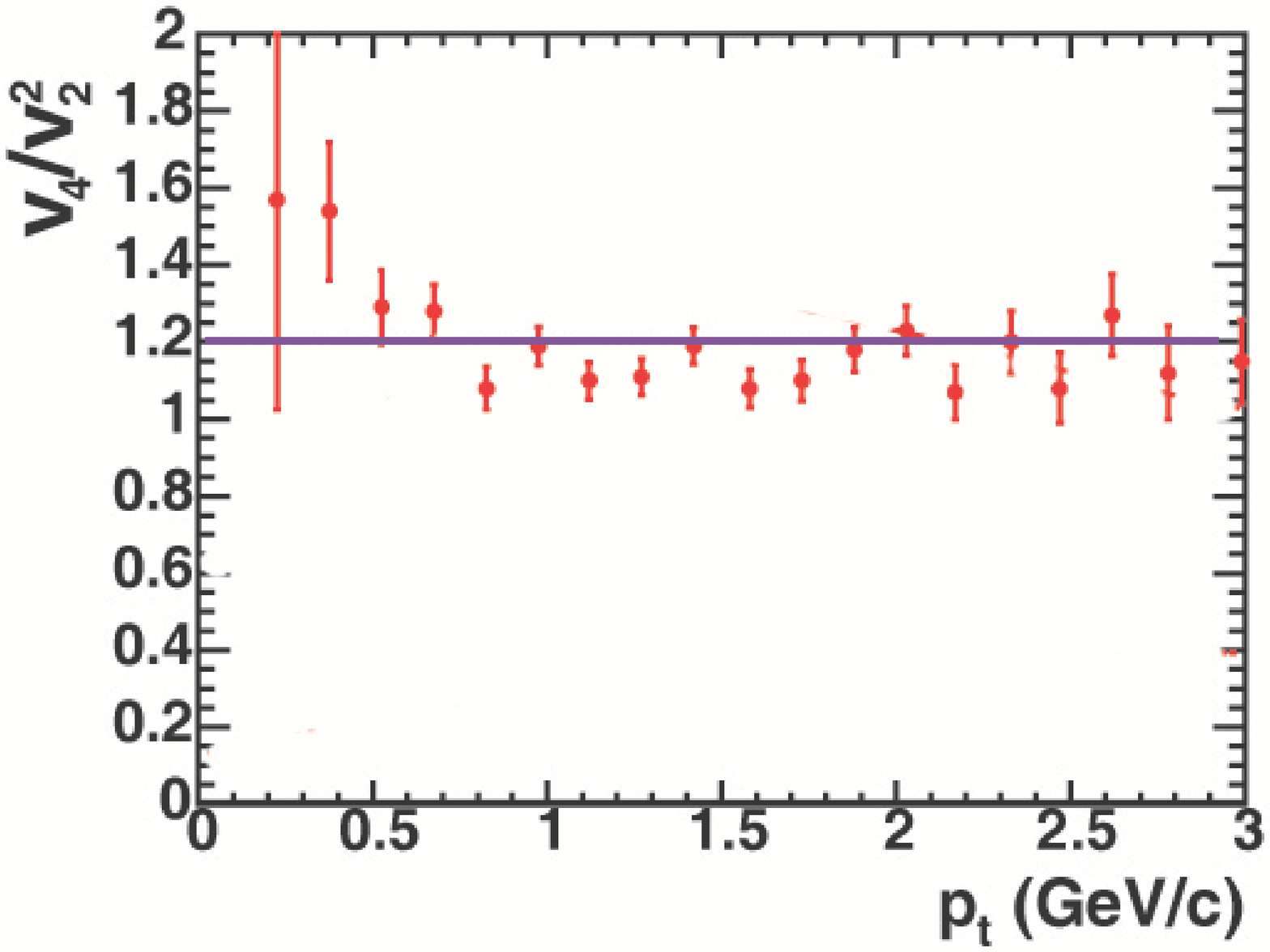}}
\caption{(Color online) Left panel (AMPT calculations): Transverse
momentum dependence of midrapidity parton anisotropic flows $v_2$,
$v_4$, and $v_6$ from minimum bias events for Au+Au at $\sqrt{s}$
= 200 A Gev with parton scattering cross section 10 mb. Also
plotted are $v_2^2$ (open triangles) and $v_2^3$ (open stars).
Figure taken from \cite{ChenLW};  Middle panel (STAR data):
Minimum bias measurements of anisotropic flow of charged hadrons
for different harmonics for Au+Au at $\sqrt{s}$ = 200 A GeV. The
dashed lines show 1.2 $v_2^2$ and 1.2$v_2^3$, respectively. Figure
taken from \cite{STAR03}; Right panel (STAR data): Ratio of
$v_4/v_2^2$ vs $p_T$  of charged hadrons for Au+Au at $\sqrt{s}$ =
200 A GeV . The  straight line represents the mean of all entries
at a value of 1.2 \cite{star2}. } \label{fig-star}
\end{figure}

The RHIC experimental data demonstrated a scaling relationship
between 2-nd flow ($v_2$) and n-th flow ($v_n$), namely
$v_{n}(p_{t})\sim v_{2}^{n/2}(p_{t})$ \cite{STAR03}. It has been
shown \cite{Kolb2,ChenLW} that such  scaling relation follows from
a naive quark coalescence model \cite{Molnari} that only allows
quarks with equal momentum to form a hadron. Denoting the meson
anisotropic flows by $v_{n,M}(p_t)$ and baryon anisotropic flows
by $v_{n,B}(p_t)$, Kolb et al. found that $v_{4,M}(p_{t}) =
(1/4)v_{2,M}^{2}(p_{t})$ for mesons and $v_{4,B}(p_{t}) =
(1/3)v_{2,B}^{2}(p_{t})$ for baryons if quarks have no
higher-order anisotropic flows. Since mesons dominate the yield of
charged particles in RHIC experimental data, the smaller scaling
factor of $1/4$ than the empirical value of about $1$ indicates
that higher-order quark anisotropic flows cannot be neglected.
Including the latter contribution, one can show that
\begin{equation}
\frac{v_{4,M}}{v_{2,M}^{2}} \approx \frac{1}{4}+\frac{1}{2}%
\frac{v_{4,q}}{v_{2,q}^{2}},\allowbreak \label{v4Mscal} ~~~~
\frac{v_{4,B}}{ v_{2,B}^2} \approx \frac{1}{3} + \frac{1}{3}
\frac{v_{4,q}}{v_{2,q}^2}\,, \label{eqn-v4}
\end{equation}
where $v_{n,q}$ denotes the quark anisotropic flows. The meson
anisotropic flows thus satisfy the scaling relations if the quark
anisotropic flows also satisfy such relations. Left panel of
Fig.~\ref{fig-star} shows the AMPT results for $v_2$, $v_4$, and
$v_6$ from minimum bias events for Au+Au at $\sqrt{s}$ = 200 A Gev
with parton scattering cross section 10 mb \cite{ChenLW}. It is
seen that the parton anisotropic flows from the AMPT model indeed
satisfy the scaling relation  $v_{n}(p_{t})\sim
v_{2}^{n/2}(p_{t})$ \cite{ChenLW}. However, as shown in the middle
panel of Fig.~\ref{fig-star}, $v_4$ is experimentally larger than
the simulation  and the ratio of $v_4/v_2^2$ is about 1.2 for
charged hadrons  \cite{STAR03,star2} (see right panel of
Fig.~\ref{fig-star}), which means that the fourth-harmonic flow of
quarks $v_4^q$ must be greater than zero. One can go one step
further and assume that the observed scaling of the hadronic $v_2$
actually results from a similar scaling occurring at the partonic
level. In this case, if one assumes \cite{Kolb2,ChenLW} that the
scaling relation for the partons is as follows:
\begin{equation}
v_4^q = (v_2^q)^2, \label{eq-parton}
 \end{equation}
 and then hadronic ratio
$v_4/v_2^2$ then equals $1/4 + 1/2 = 3/4$ for mesons and
$1/3+1/3=2/3$ for baryons, respectively.  However, since this
value is measured to be 1.2 for charged hadrons and pions, so that
either the parton scaling relation [Eq.~(\ref{eq-parton})] must
have a proportionality constant of about 2, or the simple
coalescence model needs improvement.

\section{Number-of-Nucleon Scaling of the Elliptic Flow at Intermediate Energies}

\begin{figure}
\vspace{-0.2truein}
\includegraphics[scale=0.5]{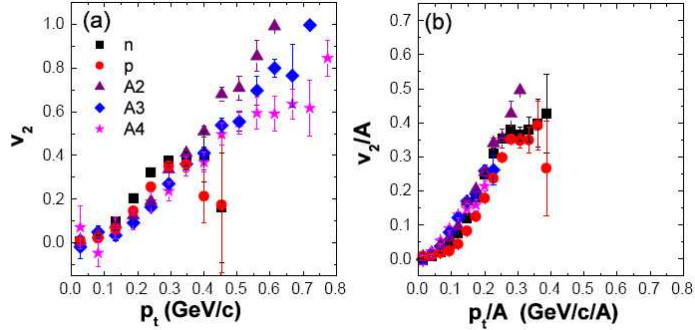}
\vspace{0.2truein} \caption{\footnotesize (a) Elliptic  flow as a
function of transverse momentum ($p_t$). Squares represent for
neutrons, circles for protons, triangles for fragments of $A$ = 2,
diamonds for $A$ = 3 and stars for $A$ = 4; (b) Elliptic flow per
nucleon as a function of transverse momentum per nucleon.  The
symbols are the same as (a).   } \label{Fig-v2-pt}
\end{figure}

In intermediate energy, we simulated $^{86}$Kr + $^{124}$Sn at 25
MeV/nucleon and impact parameter of 7 - 10 fm with IQMD
\cite{Yan}. 50,000 events have been accumulated. The system tends
to freeze-out around 120 fm/c. In this work, we extract the
following physics results at 200 fm/c. Fig.~\ref{Fig-v2-pt}(a)
shows transverse momentum dependence of elliptic flows for
mid-rapidity light fragments.  From the figure, it shows that the
elliptic flow is positive and it increases with the increasing
$p_t$. It reflects that the light clusters are preferentially
emitted within the reaction plane, and  particles with higher
transverse momentum tend to be strongly  emitted within in-plane ,
i.e. stronger positive elliptic flow. In comparison to the
elliptic flow at RHIC energies, the apparent behavior of elliptic
flow versus $p_t$ looks similar, but the mechanism is obviously
different. In intermediate energy domain, collective rotation is
one of the main mechanisms to induce the positive elliptic flow
\cite{Ritter,Peter,Shen,yg-prc,Lacey,He}. In this case, the
elliptic flow is mainly driven by the attractive mean field.
However, the strong pressure which is built in early initial
geometrical almond-type anisotropy due to the overlap zone between
both colliding nuclei in coordinate space will rapidly transforms
into the azimuthal anisotropy in momentum space at RHIC energies
\cite{J.Adams}. In other words, the elliptic flow is mainly driven
by the stronger outward pressure. Fig.~\ref{Fig-v2-pt}(b) displays
the elliptic flow per nucleon as a function of transverse momentum
per nucleon, and it looks that there exists the number of nucleon
scaling when $p_t/A < 0.25 $ GeV/$c$. This behavior is apparently
similar to the number of constituent quarks scaling of elliptic
flow versus transverse momentum per constituent quark ($p_t/n$)
for mesons and baryons which was observed at RHIC \cite{J.Adams}
and shown in Fig.~\ref{Fig-v2-pt}(b).

\begin{figure}
\vspace{-0.2truein}
\includegraphics[scale=0.5]{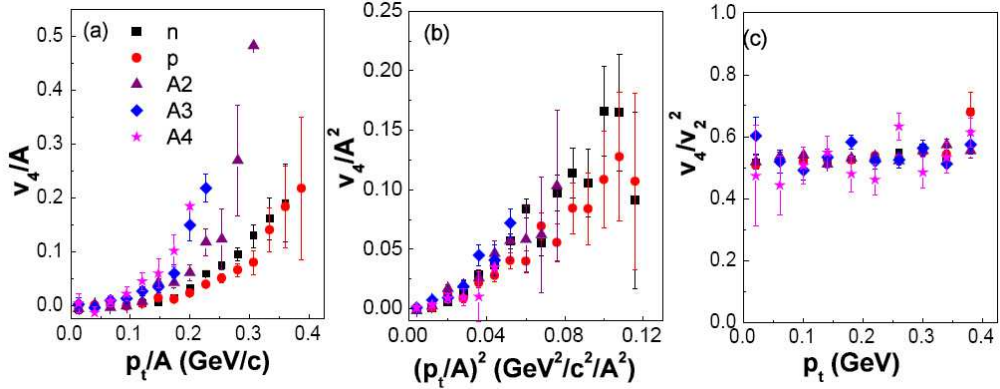}
\vspace{0.2truein} \caption{\footnotesize (a) $v_4/A$ as a
function of $p_t/A$ for different particles, namely, neutrons
(squares), protons (circles), fragments of $A$ = 2 (triangles),
$A$ = 3 (diamonds) and $A$ = 4 (stars). (b) $v_4/A^2$ as a
function of $(p_t/A)^2$. (c) the ratios of $v_4/v_2^2$ for
different particles vs  $p_t$.
 }\label{Fig-v4}
\end{figure}

Recognizing the $v_4/v_2^2$ behaviors at RHIC energies, we would
like to know what the higher order momentum anisotropy in the
intermediate energy is. So far, there is neither experimental data
nor theoretical investigation for the higher order flow, such as
$v_4$, in this energy domain. In the present work, we explore the
behavior of $v_4$ in the model calculation. Fig.~\ref{Fig-v4}
shows the feature of $v_4$. Similar to the relationship of $v_2/A$
versus $p_t/A$, we plot $v_4/A$ as a function of $p_t/A$. The
divergence of the different curves between different particles in
Fig.~\ref{Fig-v4}(a) indicates no simple scaling of nucleon number
for 4-th momentum anisotropy. However, if we plot $v_4/A^2$ versus
$(p_t/A)^2$, it looks that the points of different particles
nearly merge together and it means a certain of scaling holds
between two variables. Due to a nearly constant value of
$v_4/v_2^2$ in the studied $p_t$ range (see Fig.~\ref{Fig-v4}(c))
together with the number-of-nucleon scaling behavior of $v_2/A$ vs
$p_t/A$, $v_4/A^2$ should scale with $(p_t/A)^2$, as shown in
Fig.~\ref{Fig-v4}(b). If we assume the scaling laws of mesons
(Eq.~\ref{eqn-v4}) are also valid for A = 2 and 3 nuclear
clusters, respectively, then $v_4/v_2^2$ for A = 2 and 3 clusters
indeed give the same value of 1/2 as nucleons, as shown in
Fig.~\ref{Fig-v4}(c). Coincidentally the predicted value of the
ratio of $v_4/v_2^2$ for hadrons is also 1/2 if the matter
produced in ultra-relativistic heavy ion collisions reaches to
thermal equilibrium and its subsequent evolution follows the laws
of ideal fluid dynamics \cite{Bro}. It is interesting to note the
same ratio was predicted in two different models at very different
energies, which is of course worth to be further investigated in
near future. Overall speaking, we learn  that $v_4/v_2^2$ is
approximately 1/2 in nucleonic level coalescence mechanism, which
is different from 3/4 for mesons or 2/3 for baryons in partonic
level coalescence mechanism.

\section{Summary}

In summary, we applied AMPT and IDQMD model to investigate the
behavior of anisotropic flows, namely $v_2$ and $v_4$,  versus
transverse momentum for  hadrons in Au+Au at $\sqrt{s_{nn}}$ = 200
GeV/c as well as light nuclear fragments for 25 MeV/nucleon
$^{86}$Kr + $^{124}$Sn collisions at 7-10 fm of impact parameters.
Both $v_2$ and $v_4$ generally show positive values and increase
with $p_t$ in two very different energies. By the number scaling
of constituent-quarks or nucleons, the curves of elliptic flow for
different hadrons or nuclear fragments approximately collapse on
the similar curve, respectively, which means that there exists an
elliptic flow scaling on the constituent quark number or nucleon
number. NCQ scaling stems from  the partonic coalescence and NN
scaling originates from nucleonic coalescence. For 4-th momentum
anisotropy $v_4$, it seems to be scaled by $v_2^2$, and
$v_4/v_2^2$ is different for hadrons and nuclear fragments. AMPT
predicts a smaller value of $v_4/v_2^2$ than the data $\sim 1.2$
and the value of nuclear fragments  $\sim 0.5$. The former
indicates  that either the parton scaling relation
[Eq.~(\ref{eq-parton})] must have a proportionality constant of
about 2, or the simple quark coalescence model needs improvement.
For the later, the data is unable to be available yet. Therefore
it will be of very interesting if one can measure this ratio in
intermediate energy HIC.

\section{ Acknowledgements}
Author appreciate my collaborators and friends, especially to X.
Z. Cai,  J. H. Chen, G. L. Ma, T. Z. Yan, H. Z. Huang and W. Q.
Shen. This work was supported in part by the Shanghai Development
Foundation for Science and Technology under Grant Numbers
05XD14021 and 06JC14082, the National Natural Science Foundation
of China under Grant No 10535010, 10328259 and 10135030.

\end{document}